\documentclass[preprint]{vgtc}               
\ifpdf
  \pdfoutput=1\relax                   
  \usepackage{graphicx}                
  \DeclareGraphicsExtensions{.pdf,.png,.jpg,.jpeg} 
\else
  \usepackage{graphicx}                
  \DeclareGraphicsExtensions{.eps}     
\fi%

\graphicspath{{figures/}{pictures/}{images/}{./}} 

\usepackage{microtype}                 
\PassOptionsToPackage{warn}{textcomp}  
\usepackage{textcomp}                  
\usepackage{mathptmx}                  
\usepackage{times}                     
\usepackage{cite}                      
\usepackage{tabu}                      
\usepackage{booktabs}                  

\onlineid{2017}

\vgtccategory{Full Paper}

\vgtcinsertpkg

\preprinttext{This is the author's version, to appear in the IEEE VR 2025 conference.}

\title{Exploring Multiscale Navigation of Homogeneous and Dense Objects\\ with Progressive Refinement in Virtual Reality}

\author{Leonardo Pavanatto$^{1, 2,}$\thanks{lpavanat@vt.edu} , Alexander Giovannelli$^{1, 2}$, Brian Giera$^{2}$, Timo Bremer$^{2}$, Haichao Miao$^{2}$, Doug A. Bowman$^{1}$}

\affiliation{\scriptsize $^1$ Center for Human-Computer Interaction, Department of Computer Science, Virginia Tech, USA \\ \scriptsize $^2$ Center for Applied Scientific Computing, Lawrence Livermore National Laboratory, USA %
}

\teaser{
  \centering
  \includegraphics[width=\linewidth]{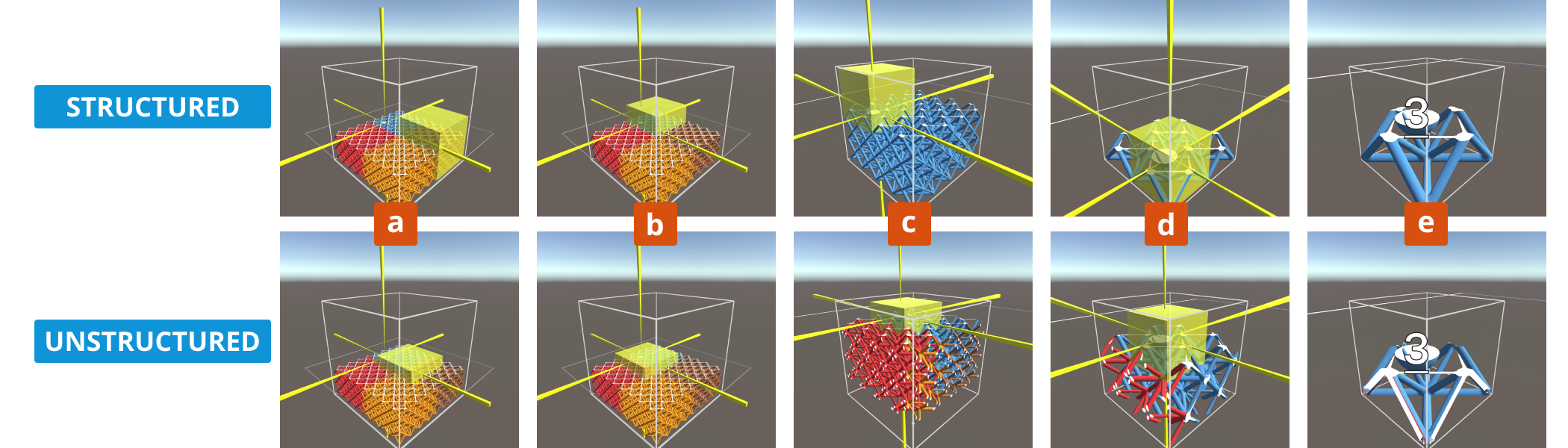}
  \caption{Navigation styles investigated in this paper. (a) in \textsc{structured}, the user can select one of eight pre-divided options; in \textsc{unstructured}, they can freely move the selection box (yellow) around; (b-d) at each level of scale, the user can select the next region to focus. \textsc{structured} selections are always precise due to the pre-division, while \textsc{unstructured} ones can include pieces of other octants (colors); (e) at the end, they reach the region of interest and can see a number.}
  \label{fig:teaser}
}
\abstract{
Locating small features in a large, dense object in virtual reality (VR) poses a significant interaction challenge. While existing multiscale techniques support transitions between various levels of scale, they are not focused on handling dense, homogeneous objects with hidden features. We propose a novel approach that applies the concept of progressive refinement to VR navigation, enabling focused inspections. We conducted a user study where we varied two independent variables in our design, navigation style (\textsc{structured} vs. \textsc{unstructured}) and display mode (\textsc{selection} vs. \textsc{everything}), to better understand their effects on efficiency and awareness during multiscale navigation. Our results showed that unstructured navigation can be faster than structured and that displaying only the selection can be faster than displaying the entire object. However, using an everything display mode can support better location awareness and object understanding.

} 

\CCScatlist{
  \CCScatTwelve{Human-centered computing}{Human computer interaction (HCI)}{Interaction paradigms}{Virtual reality};
  \CCScatTwelve{Human-centered computing}{Human computer interaction (HCI)}{Interaction paradigms}{Empirical studies in interaction design}
}

\begin{document}

\firstsection{Introduction}
\maketitle
Additive manufacturing enables the fabrication of objects with complex internal geometries \cite{chheang_virtual_2024}. While such objects can have dimensions in meters, such parts are susceptible to defects that can happen at a millimeter scale (too much or too little material across structures \cite{klacansky_virtual_2022}) and can impact the structural integrity of the part. While certain defects can be detected visually, and certain (often destructive) stress tests can be applied to the physical part, smaller defects can only be visualized through CT scans, with a manual check of each layer generated by the scan---which is impractical at scale. The scans can also be visualized in 3D, but scale differences can become a challenge during inspection processes, as it is hard to navigate through an object that is one or more orders of magnitude larger than the defects that must be visually verified. Furthermore, the dense structure of such objects implies that these defects will often be occluded inside of the object. A consequence of these challenges is that currently there are no standard procedures for validating if such products were created according to specifications. 

We investigate the use of virtual reality (VR) technologies to inspect digital twins of the fabricated parts (geometric meshes generated from either CT scans or other replication processes), enabling operators to find and assess such defects in a timely and accurate fashion while raising their understanding of how multiple defects relate and propagate across the object. Such inspections can be characterized as \textit{multiscale navigation} tasks in dense, homogeneous objects. Existing techniques have proposed multiple ways to perform multiscale navigation in VR, either continuously or discretely. These techniques, however, (1) focus mostly on structures with well-defined hierarchical meaning (such as the human body) or (2) happen in open areas (such as a map in a game). They do not account for the nuances of inspecting a dense, homogeneous object, which means defect regions may be hard to see, be hard to reach, and have reduced landmarks to support location awareness.

While in this paper we focus on the specific problem of fabricated parts, such inspections can also be relevant in other domains, such as geological analysis of soil, examining the internal defects in composite materials, evaluating the internal defects in sculptures or architectural structures, inspecting the inner workings of machinery for defects or wear, and inspecting biological structures like neural networks or cells. While all these domains still face challenges on obtaining such digital twins with high accuracy, the concepts of how to inspect them are similar to the ones discussed in this paper.

We approached this problem by applying the concept of progressive refinement to multiscale navigation. Progressive refinement \cite{kopper_rapid_2011} was proposed for selection purposes in cluttered environments, where an initial selection would include multiple objects, which would then be progressively subdivided into groups based on their proximity until a single object could be selected. While traditional navigation techniques define how to move a user through an environment, our progressive refinement navigation is object centered, where users maintain their position and the object is scaled and repositioned based on where they want to focus. While multiscale transitions can induce simulator sickness and disorientation \cite{krekhov_gullivr_2018, piumsomboon_superman_2018, cmentowski_outstanding_2019, abtahi_im_2019}, our approach allows users to select the volume on which they want to focus in a quick, systematic, and effective way.

We designed an approach called Progressive Refinement for the Inspection of Multiscale Objects (PRIMO). We identified two key characteristics of PRIMO designs: navigation style (\textsc{structured} or \textsc{unstructured}) and display mode (\textsc{selection} vs \textsc{everything}). We conducted a user study to investigate their effects on efficiency, location awareness during navigation, and overall understanding of the defects in the manufactured object. Results indicate that navigation time can be reduced when we let users select any arbitrary region while displaying only the focused subvolume. We also found evidence that displaying the entire object can aid in raising location awareness and overall object understanding. Contributions of this work include (1) the design of a technique that applies the concept of progressive refinement to the domain of multiscale navigation for the inspection of dense, homogeneous objects; and (2) a qualitative and quantitative measurement of the trade-offs of certain design choices on that design.

\section{Related Work}
\subsection{Navigating Across Levels of Scale}
Kopper et al. \cite{kopper_design_2006} introduced techniques for navigating multiscale virtual environments (MSVEs), emphasizing the need to inform users about different levels of scale and provide efficient means of transitioning between them in a \textit{discrete} fashion \cite{al_zayer_virtual_2020}. They propose navigation methods like the magnifying glass and target-based navigation. Moreover, Bacim et al. \cite{bacim_wayfinding_2009} extended this work by focusing on wayfinding aids in MSVEs, introducing techniques like a multiscale version of the world-in-miniature \cite{stoakley_virtual_1995, laviola_hands-free_2001, wingrave_overcoming_2006} and a hierarchically-structured map. Their study reveals the effectiveness of spatial and hierarchical information aids in enhancing user performance and navigation accuracy, further underlining the importance of clear wayfinding mechanisms in multiscale environments. Kouril et al. \cite{kouril_hyperlabels_2021} presented HyperLabels, a technique for navigating hierarchical molecular 3D models. HyperLabels leverages annotations and breadcrumbs to guide users through complex hierarchical structures, enhancing user comprehension and interaction efficiency.

Multiscale navigation has also been used to move through large virtual worlds. Utilizing techniques such as the ones proposed by Pierce and Pausch \cite{pierce_navigation_2004}, users can navigate through large environments by using visible landmarks as points of reference for travel. Krekhov et al. \cite{krekhov_gullivr_2018} proposed GulliVR, a technique that allows users to transition between being a giant or a regular-sized human being while walking through a large terrain. They discussed issues such as changing inter-pupillary distance (IPD) to effectively convey the perception of being a giant. They described the idea of pulling, where the system would help guide users to a point of interest. In a similar setup, Lee et al. \cite{lee_designing_2023} investigated different transitioning techniques when changing level of scale, and showed that having active control improves users' spatial awareness and performance. They also found that zooming straight out, followed by an orbital motion to reorient the user and then zooming in, presented the best spatial orientation, usability, and preference.

Another multiscale navigation approach uses continuous scaling techniques \cite{al_zayer_virtual_2020}. Such techniques gradually manipulate scale and/or speed \cite{mccrae_multiscale_2009, argelaguet_adaptive_2014} to achieve the desired movements, often automatically, with mechanisms such as viewpoint quality \cite{freitag_automatic_2016, mirhosseini_automatic_2017}, distance to surroundings \cite{ware_context_1997, mccrae_multiscale_2009, trindade_improving_2011, carvalho_dynamic_2011, cho_evaluating_2014, cho_multi-scale_2018}, and optical flow \cite{argelaguet_adaptive_2014, argelaguet_giant_2016}.

While these works define the foundation for navigating multiscale environments, the literature is lacking regarding the navigation of dense, homogeneous objects. Existing VR techniques have focused either on how to provide multiscale navigation in structures that have separable components---such as biological cells---where each level of scale has a clear meaning that supports navigation and spatial awareness; or has focused on open environments—such as zooming in and out of a map—where users don’t need to access highly occluded elements inside of a volume. When we consider objects with repeated dense structures, questions of understanding navigation and awareness become essential. Furthermore, most work in the literature has been about exploration or targeted navigation, with little discussion about how to perform systematic multiscale navigations for the purpose of inspection of marked regions of interest.

\subsection{Perception of Multiscale Navigation}
Besides mechanisms for navigation, we must also consider the effects of multiscale navigation on user perception. Piumsomboon et al. \cite{piumsomboon_superman_2018} investigated the effects of scaling a user up versus just moving in the air at normal size. They demonstrated that IPD plays a significant role in altering users' perceptions, particularly suggesting a potential coupling between IPD size and height. Similarly, Cmentowski et al. \cite{cmentowski_outstanding_2019} explored transitions between small and large levels of scale. They show the importance of smooth and fast transformations between perspectives to prevent simulator sickness and maintain spatial orientation. They further uncoupled the user from their avatar during travel mode to enhance spatial orientation and reduce reorientation efforts. Abtahi et al.  \cite{abtahi_im_2019} investigated the impact of perceived walking speed on user experience in large environments. They proposed three techniques: Ground-Level  Scaling, Eye-Level Scaling, and Seven-League Boots. Ground-Level Scaling was found to enhance user embodiment and stride length, while Seven-League Boots, although amplifying user movements, diminished positional accuracy at high gains. They further discussed avoiding scale changes on-the-fly and the use of animations. These findings underscore the importance of considering perceptual aspects in the design of multiscale navigation interfaces to enhance user experience and mitigate potential comfort issues, and the need for smooth transitions in multiscale navigation interfaces to ensure a seamless and immersive user experience.

\section{PRIMO Approach}
We designed a novel approach, Progressive Refinement for the Inspection of Multiscale Objects (PRIMO), to enable the inspection of dense, homogeneous objects. We opted to design a new approach because using traditional 2D methods: (1) the 3D nature and irregularity of the defects would require a user to go through multiple slices of data to try to understand the defects; (2) the user would likely have a lower understanding of the defect if constrained to 2D slices; and (3) it would be more difficult for the user to achieve a higher spatial awareness while navigating (by using VR we can have peripheral view as well as stereoscopy).

The initial requirements we defined included the need for a progressive approach (from global to localized visualizations) based on level of detail rendering (to render large objects but also focus on details), with reduced disorientation (to maintain smooth camera movement transitions while navigating) that achieved minimal to no simulator sickness and provided efficient navigation between levels of scale.

\subsection{Supporting Two Types of Navigation}
One of the first distinctions we made was acknowledging the difference and complementary nature between traditional navigation and multiscale navigation. While traditional navigation deals with translations and rotations across six degrees of freedom (DoF), multiscale navigation adds another three independent degrees of freedom. Managing them separately allows us to optimize them rather than trying to find a specific mapping that could control all nine DoFs together. Therefore, we opted to use a simple real-walking approach (via head tracking) for providing traditional navigation at the current level of scale. This technique is natural, intuitive to users, and less prone to simulator sickness.

For navigation across scales, we then decided to use uniform scaling with discrete levels \cite{kopper_design_2006, al_zayer_virtual_2020}. Prior research has shown that the act of scaling down into an object can produce simulator sickness \cite{krekhov_gullivr_2018, piumsomboon_superman_2018, cmentowski_outstanding_2019, abtahi_im_2019}. By defining discrete levels of scale and using simple clicks to scale up or down across levels, we could perform fast and precise multiscale traversals (half a second duration in our prototype). Such ``dashes'' have been repeatedly reported as reducing or eliminating simulator sickness \cite{krekhov_gullivr_2018, cmentowski_outstanding_2019}.

\subsection{Clipping Plane}
Another issue was determining an effective way to see the inside of a solid object. X-ray vision or other strategies that remove certain parts of the object \cite{bane_interactive_2004, lisle_clean_2022} would not be effective, as the object is homogeneous and defects may be spread through the entire volume. Our solution was to adopt a traditional clipping plane, often used in volume rendering systems \cite{viega_3d_1996, sousa_vrrrroom_2017}, such as for medical scan data and architectural modeling, to render part of the object invisible in real-time.

By allowing users to move the clipping plane with their 6-DoF hand controller, they could view any point inside the object. Initially, we provided clipping planes that could be translated or rotated arbitrarily. For the purposes of our prototype, we later decided to constrain them to 1-DoF movement along the vertical axis with no rotations, which still allowed access to any point but with a simple up-and-down movement. Users control the clipping plane through direct selection and manipulation using a simple virtual hand technique \cite{laviola_3d_2017}.

\subsection{Multiscale Navigation Style}
The next challenge was defining the levels of scale within the object and the navigation style for selecting a lower level of scale as the new focus. We decided to follow the concept of progressive refinement because of the characteristics of objects created in advanced manufacturing. As they are homogeneous and do not have clear structures to use as reference points, we needed a more systematic way of mapping the options available to users. One method is to divide the object into octants at each level of scale, giving users a simple way to keep track of the current subvolume and which subvolumes had already been inspected. Selecting an octant scales it up and makes it the new focus. We call this \textsc{structured} navigation. On the other hand, if \textsc{structured} navigation is too inflexible, we could allow users to position a cube freely inside the area of current focus, and scale up the region inside the cube to make it the new focus. We call this \textsc{unstructured} navigation. A comparison of the \textsc{structured} and \textsc{unstructured} approaches can be seen in \autoref{fig:teaser}). In both methods, the selection volume is controlled through a raycast \cite{laviola_3d_2017}.

\subsection{Display Mode}
We define the focus volume as the currently selected subvolume. To reduce occlusion and distraction, it may be beneficial to hide parts of the object outside the focus volume. This allows users to focus on their selected volume clearly, be able to see it from all sides, and easily select lower levels of scale within it. We call this the \textsc{selection} display mode. On the other hand, hiding parts of the object could affect a user's location awareness as they navigate, so we also define a second display mode called \textsc{everything}, in which the entire object (with the exception of the parts that are hidden by the clipping plane) remains visible as the user navigates through levels of scale.

\subsection{Scaling and Animation}
Instead of scaling the user down, we opted to scale the object up. The reasoning is that there are fewer modifications to camera attributes that, when not properly implemented, could further impact simulator sickness \cite{krekhov_gullivr_2018}. While this approach could reduce performance and break shading effects, our environment was constrained to a single object and we used a light source at infinity (all light rays are parallel), leading to no shadow modifications. When the user selects a lower level of scale, the system enlarges the object by a uniform scale factor of 2, animated for half a second (following suggestions from the literature \cite{krekhov_gullivr_2018, piumsomboon_superman_2018, cmentowski_outstanding_2019, abtahi_im_2019}). Thus, one octant (0.5 x 0.5 x 0.5) of the original object enlarges and takes up the same amount of space as the original object after scaling. The scale factor of 2 results from using an octet-based division (each level divided into 8 spaces). While a larger factor could have been used if we divided each level into more spaces, it could also have made selection harder. There are no guidelines in the literature regarding this issue.

\subsection{Colors at Top Level}
At the top level of scale, we decided to paint the octants with unique colors. Our objective was to provide some degree of differentiation to the user that could be general enough to be applied to objects regardless of their geometry. Such colors can guide them to better understand direction and location, helping them keep track of which octants have already been inspected. Since this type of object does not have colors, we can impose them over the object without interference. Initially, we planned to change colors on the lower levels as the user moved into lower levels of scale, but the range of colors would get smaller at each lower level and make it difficult for the user to differentiate between them---our preliminary pilots showed it could lead to more confusion.

\section{User Study}
We conducted a user study to investigate how variants of the PRIMO approach affect the efficiency, comfort, location awareness, and overall object understanding of tasks where users must navigate and inspect small regions of a dense, homogeneous object.

\begin{figure}[tb]
 \centering
 \includegraphics[width=0.8\columnwidth]{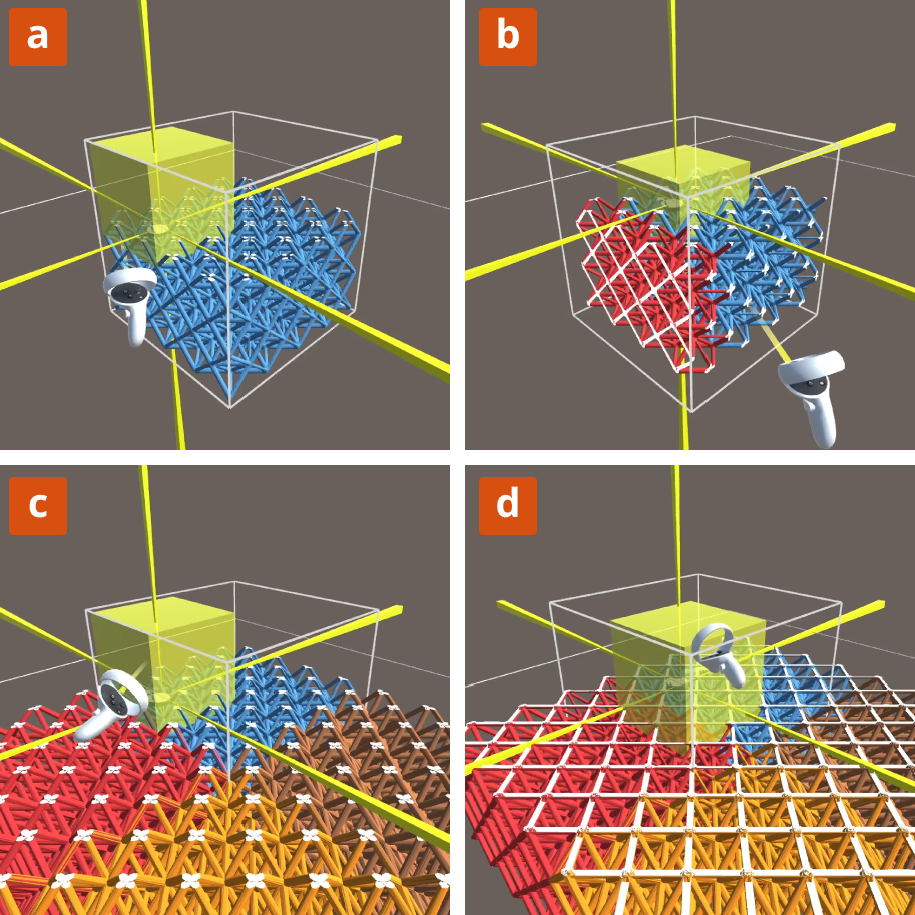}
 \caption{Conditions in this study: (a) \textsc{Selection-Structured} (subdivided showing only focus area), (b) \textsc{Selection-Unstructured} (freeform showing only focus area), (c) \textsc{Everything-Structured} (subdivided showing entire object), and (d) \textsc{Everything-Unstructured} (freeform showing entire object).}
 \label{fig:conditions}
 \vspace{-5mm}
\end{figure}

\subsection{Experimental Design}
Our study included four variations of PRIMO, by varying two independent variables with two levels each, which can be seen in \autoref{fig:conditions}. \textit{Navigation style} could be either \textsc{structured} or \textsc{unstructured}. The main rationale for studying this variable was to understand the effects of having a more systematic way of navigating versus a less constrained interaction. For \textit{display mode}, PRIMO could either display only the \textsc{selection} or \textsc{everything}. The motivation for this variable comes from the trade-off between being able to focus on only the current subvolume versus having more clutter on the screen but potentially retaining more spatial awareness. This led to four conditions: \textsc{Selection-Structured}, \textsc{Selection-Unstructured}, \textsc{Everything-Structured}, and \textsc{Everything-Unstructured}.

Both independent variables were varied within subjects, allowing all participants to complete all conditions. The presentation order was counterbalanced for each group in the following manner: trials with the same \textit{display mode} were blocked together, meaning that half of the participants completed \textsc{selection} first, and the other half completed \textsc{everything} first. Then, within the blocks, half of the participants completed \textsc{structured} navigation first, and the other half completed \textsc{unstructured} navigation first. We decided on this design to allow us to measure the effect of display mode on simulator sickness after the first two conditions.

\begin{figure}[tb]
 \centering
 \includegraphics[width=\columnwidth]{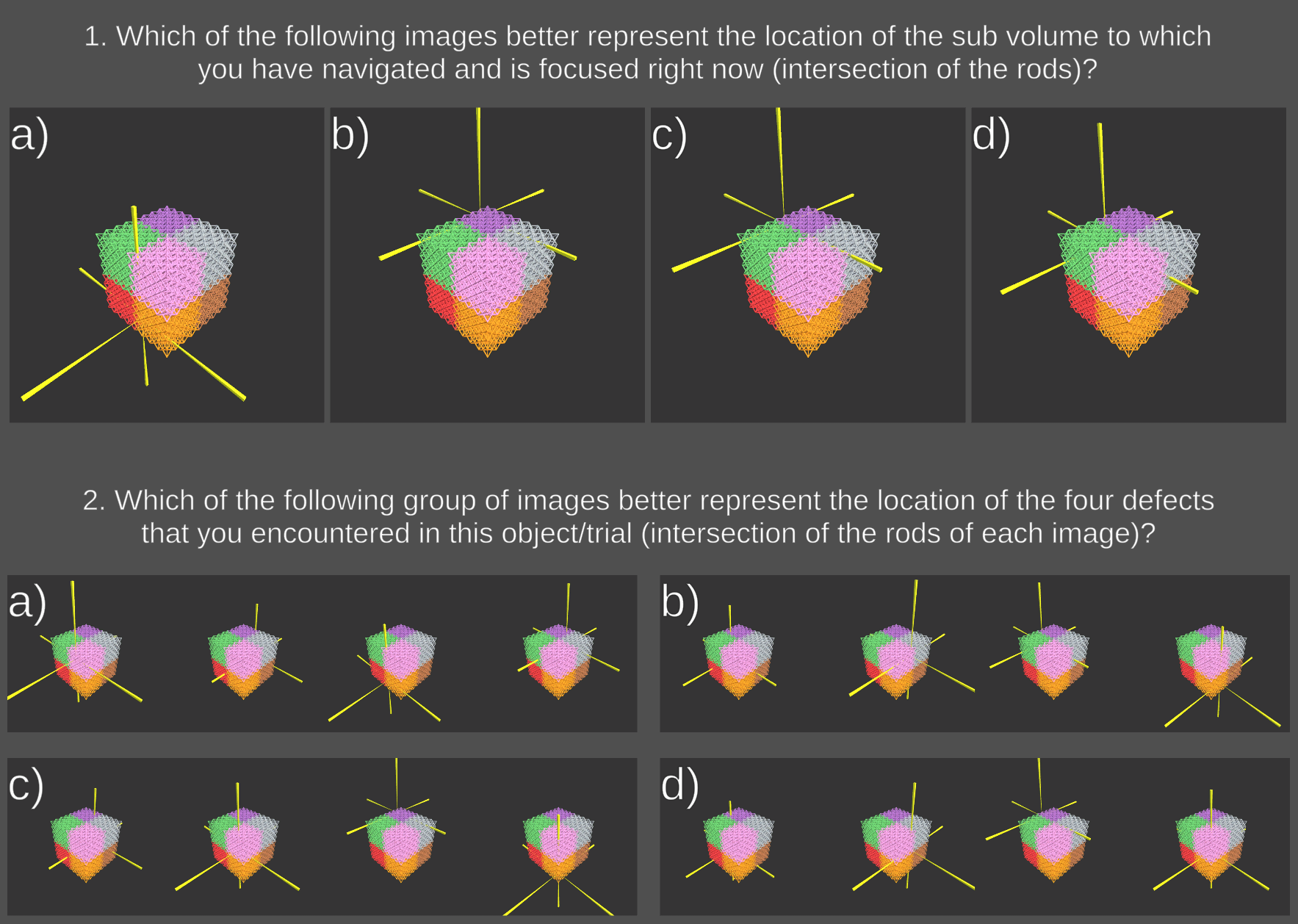}
 \caption{Questions asked inside VR: Question 1 was asked during the task and measured location awareness; question 2 was measured at the end of an object (four trials) and measured object understanding.}
 \label{fig:questions}
 \vspace{-5mm}
\end{figure}

Our objective dependent variables included the time to position the clipping plane (in milliseconds), the time to navigate from the top level of the object to a target region that required inspection (in milliseconds), and the total navigation time. We also gathered data through questionnaires that were presented inside VR. In \textbf{Question 1}, we measured the user's location awareness during navigation, allowing us to quantify how much they maintained a mental model of their focus related to the model. We designed a SAGAT-style question \cite{endsley_situation_1988} that would stop the participant in the middle of a trial and black out the environment. The question then asked: \textit{``Which of the following images better represent the location of the subvolume to which you have navigated and is focused right now (intersection of the rods)?''}, with four multiple choice answers, as seen in \autoref{fig:questions} (top). Navigation time and location awareness were measured at different times, as they might influence each other. Thus, half of the trials measured time, and half measured awareness.

After all the trials for a particular object, we were interested in overall object understanding---the ability to remember and relate the locations of the various defects within that object. \textbf{Question 2} asked: \textit{``Which of the following group of images better represent the location of the four defects that you encountered in this object/trial (intersection of the rods of each image)?''}. Again, answers were shown in multiple-choice, with each choice showing four possible defect locations (\autoref{fig:questions} (bottom)).

We also applied a simulator sickness questionnaire (SSQ). We conducted a baseline measure at the beginning of the study and a second one after the first block of two conditions was completed, to measure the differences due to the conditions. Finally, we finished the study with a semi-structured interview. We asked them about their preferred condition, which one made it easier to answer the location awareness and object understanding questions, and which one was more comfortable. We further asked them to clarify how they were feeling regarding simulator sickness.

\subsection{Hypotheses}
Our hypotheses regarding display mode and navigation style were as follows:

\textbf{H1. \textsc{Structured} will be faster than \textsc{unstructured}.}
Our argument was that \textsc{structured} had already pre-divided the object, and thus, navigation was comprised of simple point and click. Meanwhile, in \textsc{unstructured}, the user would have to position the box over the volume of interest carefully.

\textbf{H2. \textsc{Selection} will be faster than \textsc{everything}.}
We believed that \textsc{selection} would be faster than \textsc{everything} because of occlusion issues. When viewing only part of the object, the user should be able to easily see and reach the sides of the region, which are occluded if the entire object is being displayed.

\textbf{H3. \textsc{Selection} will lead to less simulator sickness than \textsc{everything}. }
As the user navigates to lower levels of scale, the focused area will appear to have the same dimensions as the original object. Thus, \textsc{selection} only displays a constant amount of the object, while \textsc{everything} occupies the user's periphery (as the rest of the object occupies previously empty space), and the scaling animation fills the field of view, which could translate into simulator sickness \cite{fernandes_combating_2016}.

\textbf{H4. \textsc{Structured} will lead to better location awareness than \textsc{unstructured}.}
While \textsc{structured} is discretized due to the object being pre-divided,  \textsc{unstructured} is a continuous selection. This implies that in a structured navigation the user may more easily remember their choices through the navigation, compared to free selections in space, which could help users maintain awareness of their location.

\textbf{H5. \textsc{Everything} will lead to better location awareness than \textsc{selection}.}
As the user moves into lower levels of scale, it may be easier for them to forget where they are or where they come from. By having \textsc{everything} display the entire object, users will have an opportunity to view their surroundings and try to understand better where exactly they are located.

\textbf{H6. \textsc{Structured} will result in a better overall object understanding than \textsc{unstructured}.}
Similarly to H4, we believed that it would be easier to retain the information about where the regions with possible defects were located by remembering the logical, discrete path to reach them.

\textbf{H7. \textsc{Everything} will result in a better overall object understanding than \textsc{selection}.}
Similarly to H5, we believe that seeing the entire object would help participants remember locations and, more importantly, the relationship between the regions with possible defects. This would happen because while navigating to one region, they would also be more likely to see the defects in other parts of the object, that are somewhat close but not exactly inside their focus area.

\subsection{Experimental Task}
\begin{figure}[tb]
 \centering
 \includegraphics[width=0.8\columnwidth]{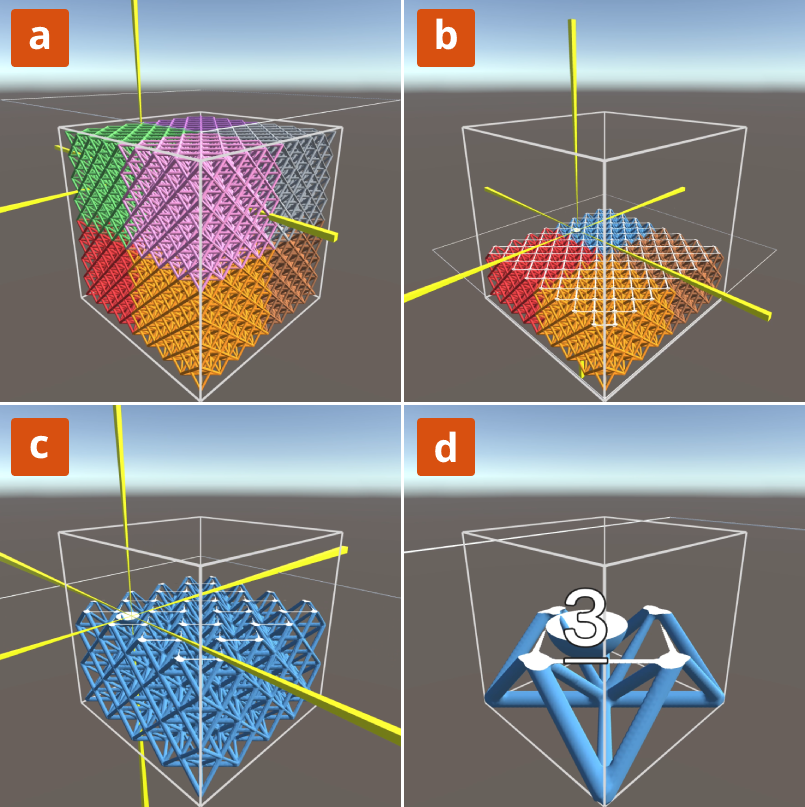}
 \caption{Steps in this study: (a) Object being inspected, with white outline showing the focus area, and yellow rods showing the location of region to inspect; (b) clipping plane cutting over the sphere at the target location; (c) navigating down into the object (multiple steps); and (d) reaching lowest level and seeing a number representing a defect.}
 \label{fig:steps}
 \vspace{-5mm}
\end{figure}

Participants had to navigate from the highest scale level of the object to the flagged region of interest at the lowest scale level. This task included four steps: (1) the participant would need to understand where the target location was---we marked this location with yellow rods parallel to each principal axis that intersected at the target, shown in \autoref{fig:steps}-a; (2) the participant would move a clipping plane to see inside of the object---the clipping plane was constrained to movements along the vertical axis and could not be rotated, and they had to move the clipping plane until it was cutting through the defect region (represented by a sphere) before they were allowed to proceed (\autoref{fig:steps}-b); (3) the participant would navigate to the target at the lowest level of scale (they were allowed to move back to a higher level of scale if they selected the wrong one) (\autoref{fig:steps}-c); (4) the participant would verbally report the number they saw in the target region (\autoref{fig:steps}-d). We decided to have a fake defect with a number instead of a real defect, because our participants were not experts in the domain of additive manufacturing, and the important thing was to demonstrate that the target region had been reached. Defects were never split between borders, to ensure that the participant could reach it.

Our rationale for the task is that while researchers can use machine learning algorithms to obtain regions where defects may exist, those are still insufficient to confirm the defects by themselves---given the difficulty and subjective evaluation of what is a defect. Further, there is an interest in understanding how such defects propagate across the object, as defects rarely occur at a single point and for no reason. Thus, our study simulates the approach of manually inspecting those pre-marked regions of interest, rather than searching from defects that could be located anywhere.

For each of the four conditions, participants inspected four objects, each containing four regions of interest to which they had to navigate. We further conducted one object inspection (also with four trials) in each condition during a training session before the main task. This led to a total of 80 trials being completed per participant: 16 training trials and 64 experiment trials. As mentioned before, half of the trials measured time and the other half measured location awareness. For each condition, the flow would be as follows: (1) training with 1 object, 2 defects (timed), and 2 defects (awareness); (2) main task with 2 objects (timed), and 2 objects (awareness). At the end of each object, we measured the overall object understanding with Question 2.

\subsection{Apparatus and Environment}
In all conditions, participants used a Meta Quest 2 head-worn display (HWD). It features a resolution of 1832 x 1920 per eye, with a refresh rate of 90Hz \footnote{https://www.meta.com/quest/products/quest-2/tech-specs}. Participants used a handheld Meta Quest 2 controller in their hand of choice. All interactions required users to point a raycast coming out of their controller and press the trigger button for selections; the button at the top of the controller allowed them to move back to a higher level of scale.

The experiment was run on a PC with an Intel i9-12900K CPU, 32GB of 3200MHz DDR4 DRAM, a Samsung SSD 980 PRO, and a GeForce RTX 3070 Ti 16GB GPU. The connection with the Meta Quest 2 was managed through the Quest link cable with up to 5Gbps transfer rate.

Participants stood in an open, unobstructed area. They could freely move around the space, only constrained by the link cable connected to the computer running the experiment. A white arrow on the floor of both the physical and virtual environments allowed users to start each trial from the same position and orientation.

\subsection{Procedure}
The Virginia Tech Institutional Review Board approved the study, which took place face-to-face at our laboratory in a single 90-minute session. We recruited participants through mailing lists and an online university recruitment system and asked them to complete a screening questionnaire for our inclusion criteria discussed in the study design section. They scheduled a session time and received a digital copy of the consent form.

Upon arrival, we greeted the participant at our laboratory. They signed the consent form and answered a background questionnaire and baseline simulator sickness questionnaire on a tablet \cite{kennedy_simulator_1993}. Next, they received general instructions through a PowerPoint presentation, and we measured and adjusted the lens position for their IPD. We followed Meta guidelines, using the 3 available IPD adjustments ($\leq61$, $61-66$, $\geq66$) \footnote{https://www.meta.com/help/quest/articles/getting-started/getting-started-with-quest-2/ipd-quest-2/}.

We asked them to stand over the white arrow and helped them wear the HWD. Once they were comfortable, we moved to the first condition, starting with the training session. Once they reached the end of the second condition (middle of the experiment), we asked them to remove the HWD, answer the simulator sickness questionnaire again, and take a mandatory five-minute break. After the break was concluded, they returned to the VR environment and completed the other two conditions. Between each condition, they were also offered an extra opportunity to take a break. Once all conditions were completed, the participant answered questions in a semi-structured interview.

\subsection{Participants}
We recruited 24 participants from the general population who fit the following inclusion criteria: (1) were at least 18 years old, (2) were proficient with the English language, (3) had normal vision (corrected or uncorrected; glasses were excluded due to equipment restrictions), and (4) had normal mobility of arms, hands, and legs (to manipulate a VR controller while walking in a space).

Twenty-four participants (aged 18 to 28; 12 male, 11 female, 1 non-binary) from the campus population took part in the experiment in individual sessions of around 90 minutes. All participants were undergraduate students from a range of disciplines. Twenty participants were right-handed, three were left-handed, and 1 was ambidextrous. Only two participants (out of three who declared themselves to be left-handed) chose to conduct the study using the left controller (the remaining left-handed participant opted to use the right controller). Twenty-two of the participants rated their fatigue level between 1 and 3 (out of 5). Twenty-one participants ranked their VR experience between 1 and 3 (out of 5), with twenty participants having used VR at least once.

\subsection{Results}
We collected our results from multiple sources. We had Google Forms that recorded the questionnaires, namely the background and simulator sickness questionnaires. From Unity, we obtained the time to complete each of the basic tasks and all the answers to the multiple-choice questions. Finally, we recorded audio files with participants' responses during the semi-structured interviews. These were transcribed by Microsoft Office Word 365, with manual verification and adjustments conducted by the authors. We conducted a statistical analysis using the \textit{JMP Pro 16} software. We used an $\alpha$ level of 0.05 in all significance tests. In the results figures, significantly different pairs are marked with * when $p\leq.05$, ** when $p\leq.01$, and *** when $p\leq.001$.

We verified normality through Shapiro-Wilk tests and normal quantile plot inspections for all cases before deciding to apply a two-way analysis of variance (ANOVA) or a non-parametric test (Wilcoxon). We performed pairwise comparisons using Tukey HSD (honestly significant difference) when appropriate. Our two factors were the display mode (\textsc{Selection} vs \textsc{Everything}) and navigation style (\textsc{Structured} vs \textsc{Unstructured}). We report partial eta squared effect sizes ($\eta_p^2$) for main effects and Cohen's $d$ for pairwise comparisons.

\subsubsection{Task Completion Time}

\begin{figure*}[tb]
 \centering
 \includegraphics[width=\linewidth]{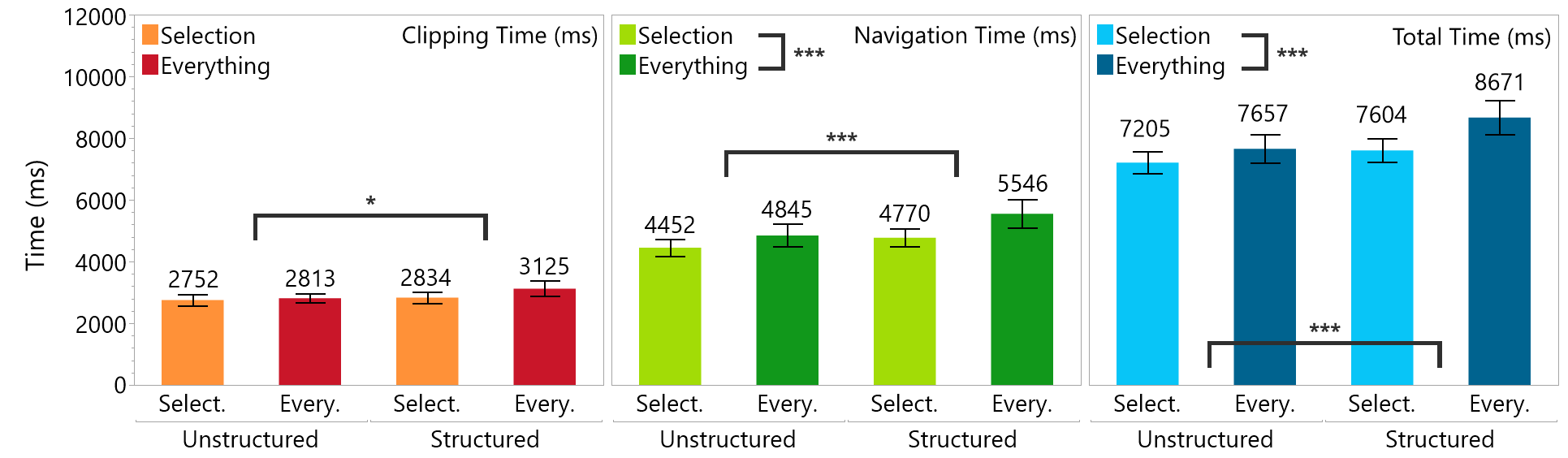}
 \caption{Time to complete: (1) clipping subtask, before starting the navigation; (2) navigation subtask, where our variables actually changed; and (3) the total task considering both. Error bars represent 95\% confidence intervals. }
 \label{fig:results}
 \vspace{-5mm}
\end{figure*}

\begin{table}[tb]
\caption{Statistics of speed to inspect measurements during the task. There were no interaction effects.}
\label{tab:speed_results}
\def\arraystretch{1.3}
\resizebox{\linewidth}{!}{%
\begin{tabular}{m{5.3em}ll}
\toprule
Measure         & Navigation Style                   & Display Mode\\
\midrule
Clipping   & $F_{1,1} = 3.99, p = 0.046; \eta_p^2 = 0.005$ & -\\
Navigation & $F_{1,1} = 8.04, p = 0.005; \eta_p^2 = 0.010$ & $F_{1,1} = 10.56, p=0.001; \eta_p^2 = 0.013$\\
Total      & $F_{1,1} = 9.78, p = 0.002; \eta_p^2 = 0.013$ & $F_{1,1} = 11.32, p \textless 0.001; \eta_p^2 = 0.015$\\
\bottomrule
\end{tabular}
}
\vspace{-6mm}
\end{table}
We measured the time to complete the task during speed trials. All the statistics in this subsection can be seen in \autoref{tab:speed_results} and represented in \autoref{fig:results}.

Regarding the \textbf{clipping time}, \textsc{Structured} ($\mu=2979.15, \sigma=1551.52$) led to a significantly larger time than \textsc{Unstructured} ($\mu=2782.51, \sigma=1151.84$), with a small effect. We did not find a significant effect of display mode or an interaction effect. Regarding the \textbf{navigation time}, \textsc{Everything} ($\mu=5195.29, \sigma=2940.08$) led to a significantly larger time than \textsc{Selection} ($\mu=4,611.26, \sigma=1969.96$), with a small effect. \textsc{Structured} ($\mu=5158.08, \sigma=2699.02$) also led to a significantly larger time than \textsc{Unstructured} ($\mu=4648.46, \sigma=2298.01$), with a small effect. We did not find an interaction effect.

Regarding the \textbf{total time}, \textsc{Everything} ($\mu=8163.99, \sigma=3646.11$) led to a significantly larger time than \textsc{Selection} ($\mu=7404.19, \sigma=2556.98$), with a small effect. \textsc{Structured} ($\mu=8137.21, \sigma=3395.41$) also led to a significantly larger time than \textsc{Unstructured} ($\mu=7430.97, \sigma=2888.39$), with a small effect. We did not find an interaction effect.

\subsubsection{Location Awareness}
We did not find any significant results for \textbf{location awareness} (Question 1), for either display mode ($F_{1,1}=0.73, p=0.39$) or navigation style ($F_{1,1}=0.05, p=0.82$), or their interaction ($F_{1,1}=0.59, p=0.59$). Averages were as follows: \textsc{Structured} ($\mu=0.68$), \textsc{Unstructured} ($\mu=0.69$), \textsc{Selection} ($\mu=0.67$), \textsc{Everything} ($\mu=0.70$).

\subsubsection{Overall Object Understanding}
Regarding the \textbf{overall object understanding} (Question 2), \textsc{Unstructured} ($\mu=0.67, \sigma=0.47$) led to a significantly larger percentage of correct answers than \textsc{Structured} ($\mu=0.56, \sigma=0.50$), with $F_{1,1}=4.88, p=0.028; \eta_p^2=0.013$ (small effect). We did not find a significant effect of display mode or an interaction effect.

\subsubsection{Simulator Sickness Questionnaire}
We subtracted the answers of the pre-exposure SSQ from the post-exposure SSQ. Pre-exposure was measured at the start of the study, and post-exposure was measured at the mandatory break in the middle of the study (after the trials with the first display mode). We then calculated the scores for nausea, oculomotor, disorientation, and total score, and performed a between-subjects Wilcoxon Signed-Rank Test. We did not find any significant effects of display mode on \textbf{simulator sickness}, for nausea ($Z=-0.36, p=0.72$), oculomotor ($Z=-0.60, p=0.55$), disorientation ($Z=0, p=1.00$), or total score ($Z=-0.47, p=0.64$). Although there was a trend for \textsc{everything} ($\mu=13.40, \sigma=19.75$) to have a higher total score than \textsc{selection} ($\mu=4.99, \sigma=9.21$), variation was too large to make this difference significant. Two participants from the \textsc{everything} condition reported feeling symptoms of motion sickness both in the questionnaire and verbally and later asked for the third optional break (between conditions 3 and 4).

\subsubsection{Qualitative Analysis}
The coding process was performed by a single experimenter using a bottom-up approach. For each transcription, the task was labeled based on the main topic being described by the participant (using Taguette \cite{rampin_taguette_2021}). New labels would be created as needed, or existing ones would be re-utilized. Once all labeling was completed, two experimenters read through them and organized the findings into themes. Those themes were then analyzed and turned into findings presented in this section. 

\paragraph{\textbf{\textsc{Everything} provides more context}}
Twenty-two participants mentioned that \textsc{everything} provided them with more context, and twenty-one said that they perceived it easier to answer Question 1 about location awareness while using it. P5 said, ``When I saw \textsc{everything} around me, it gave me a little bit more context ... I could still look over and see the colors that were nearby.'' P11 added, ``Seeing the stuff around me kept me aware of where I was within the object still. When it disappeared, I forgot exactly where I was within the object.''

Three participants also commented on how \textsc{everything} allowed them to see other defects in their peripheral view. P0 said, ``When you saw the full image, you could still kind of see where the other defects were. You can still see there is another one in the same area, as opposed to completely chopping and only having that little section.'' P17 added, ``I found pretty helpful if there were defects in the same plane ... at one point I saw there were three defects in the same plane.''

\paragraph{\textbf{\textsc{Everything} feels cluttered}}
For seven participants, \textsc{everything} ended up cluttering the environment and making them feel uncomfortable with being surrounded by the object. P15 said, ``It felt weird being inside of it.'' P8 mentioned, ``I wouldn't say it made me more disoriented, but I think it was a little more confusing when I could see everything,'' which, in turn, made it harder to select certain subvolumes: ``it was almost harder for me to pick a spot because it's a little harder to see what those subdivided regions are.'' P9 mentioned they preferred \textsc{selection} because ``then I wasn't bombarded with everything ... from a visual flood point of view was easier to ... be just focused on the one object.'' And they further discussed the issue of scaling the environment vs a single object, ``it felt like I was being pushed into the box, instead of the box disappearing ... it looks like the ground to you at the moment.'' P5 actually thought this was a good thing, ``I just enjoyed \textsc{everything} a little bit more ... looking at the structure in a very close, right in front of your face, point of view.''

\paragraph{\textbf{\textsc{Selection} helps focus on details}}
Twelve participants discussed the effects of focusing on the actual defect while using the \textsc{selection} display mode. P2 said, ``You could see more in-depth ... there wasn't anything around it, so you could see more clearly the actual sphere.'' P4 mentioned how it helped navigation, ``It was easier to focus on where I was trying to get to. It was less distracting ... having everything go away once I scaled down was powerful.'' P15, ``I didn't have the rest of the unnecessary parts all in my way.'' P20 concluded, ``In terms of finding the defect, it was a lot easier to pinpoint where exactly it was within that self-object.'' P0 further mentioned that this focus helped them identify they moved to the wrong subvolume, ``sometimes I would select it and it would look like it was in the right part, but it was just that much off.''

\paragraph{\textbf{\textsc{Structured} provides cues for location estimation}}
Eleven participants commented on how \textsc{structured} navigation provided them with cues that helped estimate their location within the object, while fourteen participants said that they perceived it easier to answer Question 1 about location awareness while using it. P4 said, ``I was specifically picking a certain section of the square, which I think made that easier.'' P3 mentioned, ``It's just very clear what color you're on, and then I'll just put that color in the back of my head and try to remember it for the end.'' P5 added that ``\textsc{structured} will kind of guide you to where you are ... I went from this section to within that section of this section.'' P11 said, ``It stuck in my memory longer when it was \textsc{structured}.'' P15 provided in-depth details, ``I remembered the color, and then if it was at the top or bottom ... then I just did that for each one.'' P23 concluded, ``if it's in subdivisions, it's easier to know where you are than if it's not.''

\paragraph{\textbf{\textsc{Unstructured} was easier and more robust}}
Ten participants discussed how \textsc{unstructured} navigation made it easier for them to select the subvolume they wanted. P10 said, ``I felt like I had more control over where I was going. It just felt easier for me to navigate.'' P3 explained, ``I felt I could just immediately scale down right into the point I was looking at.'' P4 added, ``when I scaled down, the region would be in the center.'' P7 made a point about occlusion making it harder for \textsc{structured}, ``With \textsc{structured} you had to point at a specific angle, either point higher or lower ... but for the \textsc{unstructured} I could just point and then click wherever I'm comfortable, which made it a lot easier.''

Four participants mentioned that \textsc{unstructured} was less prone to wrong volume selections than \textsc{structured}. P5 said, ``there were times when I would be pointing using \textsc{structured} and it would just blip over to the section nearby. Whereas when you're \textsc{unstructured} ... it still can contain the defect.'' P10 added, ``with the \textsc{structured} ones, it was more you either hit it or you don't. With the other ones it was more accurate, felt I had more control over the system.'' P14, ``With the one where I could move the box on my own, I got to pick in-between colors.''

\paragraph{\textbf{\textsc{Structured} requires less ray-casting precision}}
Eight participants also described that \textsc{structured} required less precision from them during the process of pointing their raycast. P2 said, ``There wasn't as much guessing of where to put the pointer. It was more guided.'' P4 added, ``I just pointed to the specific square I needed to go.'' P9 expanded, ``I don't have to worry about really getting it ... as long as it's in that area, I'm fine.'' P13 claimed, ``It's hard to position the yellow cube [in \textsc{unstructured}] exactly where you want it.'' P17 said, ``It was a little bit difficult to pinpoint the defect [in \textsc{unstructured}]. Versus if you have \textsc{structured} you can sort of just highlight a certain area.''

\paragraph{\textbf{Simulator sickness was minimal}}
Twenty participants reported no simulator sickness at all, while two reported mild symptoms of simulator sickness. P6 said, ``I feel as the experiment continued the dizziness started to kick in a little bit more, but maybe it's because of fatigue.'' P15 said, ``\textsc{everything unstructured} was probably the most dizzying because it was a lot of movement and there was a lot around me ... moving myself and I'm inside of something, it was just a lot of stimulus going on all at once.''

\section{Discussion}
\paragraph{\textbf{Performance}}
Our first hypothesis stated that \textsc{structured} would be faster than \textsc{unstructured} (\textbf{H1}). Evidence suggests that this hypothesis is false. On the contrary, we found that during navigation, \textsc{unstructured} was 10\% faster than \textsc{structured} on average. Based on the qualitative data, the reason was clear: while in \textsc{structured} participants do have the advantage of already having the object pre-divided into subvolumes and only having to select which one they want, they need to reposition their ray at every level of scale to select the next subvolume. With \textsc{unstructured}, on the other hand, the selection box would show at the intersection between the participant's raycast and the object. This means that if the participant focused on the region of interest at the top level, they would only have to press the confirmation button at each subsequent level of scale. Therefore, even if the original placement of the selection box took longer, they could more quickly navigate down to the targeted location. Obviously, these results might be different if the region of interest was not shown to the user from the top level, and they had to search for it instead. Unexpectedly, although small, we found a difference during the clipping portion of the task, with \textsc{structured} being slower. We could speculate that this is due to the context-switching time or fatigue being influenced by the conditions, but we cannot affirm this with our current data.

We also hypothesized that the \textsc{selection} display mode would be faster than \textsc{everything} (\textbf{H2}). Our results support this hypothesis. During navigation, \textsc{selection} display mode was 12\% faster than \textsc{everything} on average. Based on qualitative data, we can attribute this to a couple of reasons. First, \textsc{selection} only showed the focused object, and thus, participants could clearly see and focus on selecting the next subvolume. In \textsc{everything}, however, the other regions outside of the focused/selected volume still caused occlusion. Not only could participants not see the sides of the focused volume, but they would also have to orient their hands awkwardly or physically walk around to get the ray in the right place. This was even stronger when using \textsc{structured} navigation, as the raycast selected the first subdivision it touched, while with \textsc{unstructured} participants could still move their hands in depth to move the selection box in depth. Second, from a pure visualization point of view, displaying the entire object while only allowing for further navigation inside of the focus region made participants confused in cases where they selected the wrong level of scale, but after navigating they could still see the marker for the correct region of interest. In some cases it took them a couple of seconds to realize the mistake before taking action to move back to the higher level and select the other subvolume.

\paragraph{\textbf{Simulator Sickness}}
We hypothesized that \textsc{selection} would lead to less simulator sickness than \textsc{everything} (\textbf{H3}). Our objective results do not support this hypothesis. There was a trend for \textsc{everything} to score higher than \textsc{selection}, but this did not lead to a significant difference due to a high variance in the measure. Only two out of twenty-four participants complained of mild simulator sickness. This is a positive result for the techniques, as each participant completed a large number of trials and still experienced only mild levels of sickness. In the qualitative results, none of the participants commented on a clear difference between the techniques in regards to sickness, dizziness, or disorientation. A few participants did note, however, that the \textsc{everything} display mode created a zooming-in effect with a lot happening visually, where the ``entire world'' was getting larger around them as they shrunk, as opposed to the \textsc{selection} display mode where the object felt like it was getting larger. 

\paragraph{\textbf{Location Awareness}}
Our fourth hypothesis was that \textsc{structured} would lead to better location awareness than \textsc{unstructured} (\textbf{H4}). However, we found no differences between the conditions on the accuracy of answering question 1. From interviews, we got mixed results. Participants who believed \textsc{unstructure} was better argued that they only needed to focus on their target region of interest and then get there faster, having to remember only that one point. Participants who believed \textsc{structured} was better argued that (1) remembering the path they took to get there was easier (because they just had to think about the discrete selection steps, and (2) since the object was already pre-divided, they could think of the octants when trying to pinpoint the focused one. Our lack of significant results may have been due to participants using different strategies to maintain awareness of their location during navigation.

In the fifth hypothesis, we argued that \textsc{everything} would lead to better location awareness than \textsc{selection} (\textbf{H5}). Again, the data for location awareness accuracy do not support this hypothesis. However, our interview data revealed that users perceived it to be true. More than 87\% of participants mentioned one or both of the \textsc{everything} techniques as providing better location awareness. In their comments, they extensively mentioned how using the \textsc{selection} display mode led to them trying to memorize where they were going or which steps they took as they navigated. Some memorization was still needed in the \textsc{everything} mode, but participants could also use the peripheral view of their location in the context of the whole object as they navigated. This suggests that everything may have some benefits for location awareness, but we were not able to measure it objectively because the task was not complicated enough for the memorization used in the \textsc{selection} display mode to be detrimental. We suggest that a future study with a larger number of levels of scale or with greater scale differences between those levels could revisit this hypothesis for confirmation.

\paragraph{\textbf{Overall Object Understanding}}
We hypothesized that \textsc{structured} would provide a better understanding of all defect locations than \textsc{unstructured} (\textbf{H6}). Our objective results do not support this hypothesis. We found that \textsc{unstructured} was 17\% more accurate than \textsc{structured} on question 2, which asked for the locations of all four defects in an object. Although qualitative results showed that participants believed that structuring the navigation process led to a more concise and organized amount of information to remember and combine afterward, the objective measure indicates that the act of placing the selection box in \textsc{unstructured} actually gave people a better cue to remember the defect locations.

Finally, the last hypothesis was that \textsc{everything} would result in a better understanding of all defect locations than \textsc{selection} (\textbf{H7}). Objective measures did not find a difference between the display modes. From the interviews, however, participants mentioned that \textsc{everything} was advantageous because, in some scenarios, they could partially see one defect from a distance while navigating to a different defect region, and that would refresh the location relationship between the defects in their minds. Similarly to H5, we suggest that this should be revisited with a more complicated task, especially one that allows users to navigate from one defect to the next one instead of going back all the way to the top to start the next trial. Alternatively, the task could have more than four defects in each object, leading to higher chances of seeing other defects while making it harder to remember all the defect locations.

\paragraph{Implications} Our findings bring some implications to the domain: in VR, we should give people unstructured ways to navigate when they know where they are going; on the other hand, a structured navigation could still be used for systematic search in an object where the location of defects is unknown, although another study should investigate that; we should prefer to display the entire object to enhance spatial awareness, but we should hide peripheral portions of the object when selecting the next piece to navigate.

\section{Limitations}
This work includes some limitations. (1) While we designed our study scenario with stakeholders who work on these tasks daily, a follow-up study should be performed in a real-world task and environment. (2) As we compared variations of PRIMO, we lacked a direct comparison with traditional techniques. (3) Our findings are based on a task that reduces wayfinding and focuses on travel, and results could be different if the location of potential defects were unknown. Our rationale for this choice was that machine learning can be used to narrow down regions of interest, as navigating through the entire object could be impractical due to time constraints.

Finally, (4) we limited our study to a single cuboid object and did not consider variability in the structure in the study. Our approach does expand to other form factors, as objects can still be broken into smaller cuboids. The main question would pertain to objects where one of the dimensions is much larger than the others; the technique might need to be adapted for such asymmetries, and a further study focusing on designing those adaptations should be conducted.

\section{Conclusions and Future Work}
In this paper, we explored the design space of VR multiscale navigation techniques in dense, homogeneous objects while using the concept of progressive refinement. We proposed PRIMO, an approach that allows users to traverse objects, such as those created through advanced manufacturing, to inspect small regions that have been flagged as potentially having defects. Through a user study, we varied two independent variables of the interaction design to understand the trade-offs and obtain guidelines that could be generalized for guiding practitioners interested in applying multiscale navigation for the inspection of objects that are dense and homogeneous.

Our results showed that navigation time can be minimized by allowing the user to select any arbitrary region within the object and by displaying only the currently focused subvolume. We also found evidence that unstructured navigation can lead to a better overall object understanding than structured navigation. We found qualitative results suggesting that displaying everything may also help with location awareness and overall object understanding, though we were not able to corroborate those with objective data. Based on these results, and giving more weight to spatial awareness than speed, we suggest that an optimal hybrid technique for this domain would use unstructured navigation and would display only the selected subvolume by default, but with the ability to toggle the display of everything.

For future work, we plan to delve deeper into the wayfinding cues that can support users in better understanding the navigations they took at each step. Some candidates include using spatial breadcrumbs to display representations of the previous models the user has been to, spatial trees to organize the division of the object visually, and ghost representations of the hidden objects that could be accessed when needed.

\acknowledgments{
This work was supported by the U.S. DOE LLNL-LDRD 23-SI-003 and was performed under the auspices of the U.S. Department of Energy by Lawrence Livermore National Laboratory under Contract DE-AC52-07NA27344 (LLNL-CONF-863666).
}

\bibliographystyle{abbrv-doi}

\bibliography{references}

\begin{thebibliography}{10}

\bibitem{abtahi_im_2019}
P.~Abtahi, M.~Gonzalez-Franco, E.~Ofek, and A.~Steed.
\newblock I'm a {Giant}: {Walking} in {Large} {Virtual} {Environments} at {High} {Speed} {Gains}.
\newblock In {\em Proceedings of the 2019 {CHI} {Conference} on {Human} {Factors} in {Computing} {Systems}}, pp. 1--13. ACM, Glasgow Scotland Uk, May 2019. doi: {{%
10\hspace{.1pt}\discretionary{.}{%
}{.}\hspace{.4pt}1145\discretionary{/}{%
}{/}3290605\hspace{.1pt}\discretionary{.}{%
}{.}\hspace{.4pt}3300752}}


\bibitem{al_zayer_virtual_2020}
M.~Al~Zayer, P.~MacNeilage, and E.~Folmer.
\newblock Virtual {Locomotion}: {A} {Survey}.
\newblock {\em IEEE Transactions on Visualization and Computer Graphics}, 26(6):2315--2334, June 2020. doi: {{%
10\hspace{.1pt}\discretionary{.}{%
}{.}\hspace{.4pt}1109\discretionary{/}{%
}{/}TVCG\hspace{.1pt}\discretionary{.}{%
}{.}\hspace{.4pt}2018\hspace{.1pt}\discretionary{.}{%
}{.}\hspace{.4pt}2887379}}


\bibitem{argelaguet_adaptive_2014}
F.~Argelaguet.
\newblock Adaptive navigation for virtual environments.
\newblock In {\em 2014 {IEEE} {Symposium} on {3D} {User} {Interfaces} ({3DUI})}, pp. 123--126. IEEE, Minneapolis, MN, Mar. 2014. doi: {{%
10\hspace{.1pt}\discretionary{.}{%
}{.}\hspace{.4pt}1109\discretionary{/}{%
}{/}3DUI\hspace{.1pt}\discretionary{.}{%
}{.}\hspace{.4pt}2014\hspace{.1pt}\discretionary{.}{%
}{.}\hspace{.4pt}7027325}}


\bibitem{argelaguet_giant_2016}
F.~Argelaguet and M.~Maignant.
\newblock {GiAnt}: stereoscopic-compliant multi-scale navigation in {VEs}.
\newblock In {\em Proceedings of the 22nd {ACM} {Conference} on {Virtual} {Reality} {Software} and {Technology}}, pp. 269--277. ACM, Munich Germany, Nov. 2016. doi: {{%
10\hspace{.1pt}\discretionary{.}{%
}{.}\hspace{.4pt}1145\discretionary{/}{%
}{/}2993369\hspace{.1pt}\discretionary{.}{%
}{.}\hspace{.4pt}2993391}}


\bibitem{bacim_wayfinding_2009}
F.~Bacim, D.~Bowman, and M.~Pinho.
\newblock Wayfinding techniques for {multiScale} virtual environments.
\newblock In {\em 2009 {IEEE} {Symposium} on {3D} {User} {Interfaces}}, pp. 67--74. IEEE, Lafayette, LA, USA, 2009. doi: {{%
10\hspace{.1pt}\discretionary{.}{%
}{.}\hspace{.4pt}1109\discretionary{/}{%
}{/}3DUI\hspace{.1pt}\discretionary{.}{%
}{.}\hspace{.4pt}2009\hspace{.1pt}\discretionary{.}{%
}{.}\hspace{.4pt}4811207}}


\bibitem{bane_interactive_2004}
R.~Bane and T.~Hollerer.
\newblock Interactive tools for virtual x-ray vision in mobile augmented reality.
\newblock In {\em Third {IEEE} and {ACM} {International} {Symposium} on {Mixed} and {Augmented} {Reality}}, pp. 231--239, 2004. doi: {{%
10\hspace{.1pt}\discretionary{.}{%
}{.}\hspace{.4pt}1109\discretionary{/}{%
}{/}ISMAR\hspace{.1pt}\discretionary{.}{%
}{.}\hspace{.4pt}2004\hspace{.1pt}\discretionary{.}{%
}{.}\hspace{.4pt}36}}


\bibitem{carvalho_dynamic_2011}
F.~Carvalho, D.~R. Trindade, P.~F. Dam, A.~Raposo, and I.~H. F.~d. Santos.
\newblock Dynamic {Adjustment} of {Stereo} {Parameters} for {Virtual} {Reality} {Tools}.
\newblock In {\em 2011 {XIII} {Symposium} on {Virtual} {Reality}}, pp. 66--72. IEEE, Uberlandia, TBD, Brazil, May 2011. doi: {{%
10\hspace{.1pt}\discretionary{.}{%
}{.}\hspace{.4pt}1109\discretionary{/}{%
}{/}SVR\hspace{.1pt}\discretionary{.}{%
}{.}\hspace{.4pt}2011\hspace{.1pt}\discretionary{.}{%
}{.}\hspace{.4pt}30}}


\bibitem{chheang_virtual_2024}
V.~Chheang, B.~T. Weston, R.~W. Cerda, B.~Au, B.~Giera, P.-T. Bremer, and H.~Miao.
\newblock A {Virtual} {Environment} for {Collaborative} {Inspection} in {Additive} {Manufacturing}, 2024.

\bibitem{cho_evaluating_2014}
I.~Cho, J.~Li, and Z.~Wartell.
\newblock Evaluating dynamic-adjustment of stereo view parameters in a multi-scale virtual environment.
\newblock In {\em 2014 {IEEE} {Symposium} on {3D} {User} {Interfaces} ({3DUI})}, pp. 91--98. IEEE, MN, USA, Mar. 2014. doi: {{%
10\hspace{.1pt}\discretionary{.}{%
}{.}\hspace{.4pt}1109\discretionary{/}{%
}{/}3DUI\hspace{.1pt}\discretionary{.}{%
}{.}\hspace{.4pt}2014\hspace{.1pt}\discretionary{.}{%
}{.}\hspace{.4pt}6798848}}


\bibitem{cho_multi-scale_2018}
I.~Cho, J.~Li, and Z.~Wartell.
\newblock Multi-{Scale} {7DOF} {View} {Adjustment}.
\newblock {\em IEEE Transactions on Visualization and Computer Graphics}, 24(3):1331--1344, Mar. 2018. doi: {{%
10\hspace{.1pt}\discretionary{.}{%
}{.}\hspace{.4pt}1109\discretionary{/}{%
}{/}TVCG\hspace{.1pt}\discretionary{.}{%
}{.}\hspace{.4pt}2017\hspace{.1pt}\discretionary{.}{%
}{.}\hspace{.4pt}2668405}}


\bibitem{cmentowski_outstanding_2019}
S.~Cmentowski, A.~Krekhov, and J.~Krüger.
\newblock Outstanding: {A} {Multi}-{Perspective} {Travel} {Approach} for {Virtual} {Reality} {Games}.
\newblock In {\em Proceedings of the {Annual} {Symposium} on {Computer}-{Human} {Interaction} in {Play}}, pp. 287--299. ACM, Barcelona Spain, Oct. 2019. doi: {{%
10\hspace{.1pt}\discretionary{.}{%
}{.}\hspace{.4pt}1145\discretionary{/}{%
}{/}3311350\hspace{.1pt}\discretionary{.}{%
}{.}\hspace{.4pt}3347183}}


\bibitem{endsley_situation_1988}
M.~R. Endsley.
\newblock Situation awareness global assessment technique ({SAGAT}).
\newblock In {\em Proceedings of the {IEEE} 1988 {National} {Aerospace} and {Electronics} {Conference}}, pp. 789--795 vol.3, May 1988. doi: {{%
10\hspace{.1pt}\discretionary{.}{%
}{.}\hspace{.4pt}1109\discretionary{/}{%
}{/}NAECON\hspace{.1pt}\discretionary{.}{%
}{.}\hspace{.4pt}1988\hspace{.1pt}\discretionary{.}{%
}{.}\hspace{.4pt}195097}}


\bibitem{fernandes_combating_2016}
A.~S. Fernandes and S.~K. Feiner.
\newblock Combating {VR} sickness through subtle dynamic field-of-view modification.
\newblock In {\em 2016 {IEEE} {Symposium} on {3D} {User} {Interfaces} ({3DUI})}, pp. 201--210, 2016. doi: {{%
10\hspace{.1pt}\discretionary{.}{%
}{.}\hspace{.4pt}1109\discretionary{/}{%
}{/}3DUI\hspace{.1pt}\discretionary{.}{%
}{.}\hspace{.4pt}2016\hspace{.1pt}\discretionary{.}{%
}{.}\hspace{.4pt}7460053}}


\bibitem{freitag_automatic_2016}
S.~Freitag, B.~Weyers, and T.~W. Kuhlen.
\newblock Automatic speed adjustment for travel through immersive virtual environments based on viewpoint quality.
\newblock In {\em 2016 {IEEE} {Symposium} on {3D} {User} {Interfaces} ({3DUI})}, pp. 67--70. IEEE, Greenville, SC, USA, Mar. 2016. doi: {{%
10\hspace{.1pt}\discretionary{.}{%
}{.}\hspace{.4pt}1109\discretionary{/}{%
}{/}3DUI\hspace{.1pt}\discretionary{.}{%
}{.}\hspace{.4pt}2016\hspace{.1pt}\discretionary{.}{%
}{.}\hspace{.4pt}7460033}}


\bibitem{kennedy_simulator_1993}
R.~S. Kennedy, N.~E. Lane, K.~S. Berbaum, and M.~G. Lilienthal.
\newblock Simulator {Sickness} {Questionnaire}: {An} {Enhanced} {Method} for {Quantifying} {Simulator} {Sickness}.
\newblock {\em The International Journal of Aviation Psychology}, 3(3):203--220, July 1993. doi: {{%
10\hspace{.1pt}\discretionary{.}{%
}{.}\hspace{.4pt}1207\discretionary{/}{%
}{/}s15327108ijap0303\_3}}


\bibitem{klacansky_virtual_2022}
P.~Klacansky, H.~Miao, A.~Gyulassy, A.~Townsend, K.~Champley, J.~Tringe, V.~Pascucci, and P.-T. Bremer.
\newblock Virtual {Inspection} of {Additively} {Manufactured} {Parts}.
\newblock In {\em 2022 {IEEE} 15th {Pacific} {Visualization} {Symposium} ({PacificVis})}, pp. 81--90, 2022. doi: {{%
10\hspace{.1pt}\discretionary{.}{%
}{.}\hspace{.4pt}1109\discretionary{/}{%
}{/}PacificVis53943\hspace{.1pt}\discretionary{.}{%
}{.}\hspace{.4pt}2022\hspace{.1pt}\discretionary{.}{%
}{.}\hspace{.4pt}00017}}


\bibitem{kopper_rapid_2011}
R.~Kopper, F.~Bacim, and D.~A. Bowman.
\newblock Rapid and accurate {3D} selection by progressive refinement.
\newblock In {\em 2011 {IEEE} {Symposium} on {3D} {User} {Interfaces} ({3DUI})}, pp. 67--74. IEEE, Singapore, Singapore, Mar. 2011. doi: {{%
10\hspace{.1pt}\discretionary{.}{%
}{.}\hspace{.4pt}1109\discretionary{/}{%
}{/}3DUI\hspace{.1pt}\discretionary{.}{%
}{.}\hspace{.4pt}2011\hspace{.1pt}\discretionary{.}{%
}{.}\hspace{.4pt}5759219}}


\bibitem{kopper_design_2006}
R.~Kopper, {Tao Ni}, D.~Bowman, and M.~Pinho.
\newblock Design and {Evaluation} of {Navigation} {Techniques} for {Multiscale} {Virtual} {Environments}.
\newblock In {\em {IEEE} {Virtual} {Reality} {Conference} ({VR} 2006)}, pp. 175--182. IEEE, Alexandria, VA, USA, 2006. doi: {{%
10\hspace{.1pt}\discretionary{.}{%
}{.}\hspace{.4pt}1109\discretionary{/}{%
}{/}VR\hspace{.1pt}\discretionary{.}{%
}{.}\hspace{.4pt}2006\hspace{.1pt}\discretionary{.}{%
}{.}\hspace{.4pt}47}}


\bibitem{kouril_hyperlabels_2021}
D.~Kouril, T.~Isenberg, B.~Kozlikova, M.~Meyer, M.~E. Groller, and I.~Viola.
\newblock {HyperLabels}: {Browsing} of {Dense} and {Hierarchical} {Molecular} {3D} {Models}.
\newblock {\em IEEE Transactions on Visualization and Computer Graphics}, 27(8):3493--3504, Aug. 2021. doi: {{%
10\hspace{.1pt}\discretionary{.}{%
}{.}\hspace{.4pt}1109\discretionary{/}{%
}{/}TVCG\hspace{.1pt}\discretionary{.}{%
}{.}\hspace{.4pt}2020\hspace{.1pt}\discretionary{.}{%
}{.}\hspace{.4pt}2975583}}


\bibitem{krekhov_gullivr_2018}
A.~Krekhov, S.~Cmentowski, K.~Emmerich, M.~Masuch, and J.~Krüger.
\newblock {GulliVR}: {A} {Walking}-{Oriented} {Technique} for {Navigation} in {Virtual} {Reality} {Games} {Based} on {Virtual} {Body} {Resizing}.
\newblock In {\em Proceedings of the 2018 {Annual} {Symposium} on {Computer}-{Human} {Interaction} in {Play}}, pp. 243--256. ACM, Melbourne VIC Australia, Oct. 2018. doi: {{%
10\hspace{.1pt}\discretionary{.}{%
}{.}\hspace{.4pt}1145\discretionary{/}{%
}{/}3242671\hspace{.1pt}\discretionary{.}{%
}{.}\hspace{.4pt}3242704}}


\bibitem{laviola_hands-free_2001}
J.~J. LaViola, D.~A. Feliz, D.~F. Keefe, and R.~C. Zeleznik.
\newblock Hands-free multi-scale navigation in virtual environments.
\newblock In {\em Proceedings of the 2001 symposium on {Interactive} {3D} graphics}, pp. 9--15. ACM, Mar. 2001. doi: {{%
10\hspace{.1pt}\discretionary{.}{%
}{.}\hspace{.4pt}1145\discretionary{/}{%
}{/}364338\hspace{.1pt}\discretionary{.}{%
}{.}\hspace{.4pt}364339}}


\bibitem{laviola_3d_2017}
J.~J. LaViola, E.~Kruijff, R.~P. McMahan, D.~A. Bowman, and I.~P. Poupyrev.
\newblock {\em {3D} {User} {Interfaces}: {Theory} and {Practice}}.
\newblock Addison-Wesley Professional, 2nd ed., 2017.

\bibitem{lee_designing_2023}
J.-I. Lee, P.~Asente, and W.~Stuerzlinger.
\newblock Designing {Viewpoint} {Transition} {Techniques} in {Multiscale} {Virtual} {Environments}.
\newblock In {\em 2023 {IEEE} {Conference} {Virtual} {Reality} and {3D} {User} {Interfaces} ({VR})}, pp. 680--690. IEEE, Shanghai, China, Mar. 2023. doi: {{%
10\hspace{.1pt}\discretionary{.}{%
}{.}\hspace{.4pt}1109\discretionary{/}{%
}{/}VR55154\hspace{.1pt}\discretionary{.}{%
}{.}\hspace{.4pt}2023\hspace{.1pt}\discretionary{.}{%
}{.}\hspace{.4pt}00083}}


\bibitem{lisle_clean_2022}
L.~Lisle, F.~Lu, S.~Davari, I.~Tahmid, A.~Giovannelli, C.~Llo, L.~Pavanatto, L.~Zhang, L.~Schlueter, and D.~A. Bowman.
\newblock Clean the {Ocean}: {An} {Immersive} {VR} {Experience} {Proposing} {New} {Modifications} to {Go}-{Go} and {WiM} {Techniques}.
\newblock In {\em 2022 {IEEE} {Conference} on {Virtual} {Reality} and {3D} {User} {Interfaces} {Abstracts} and {Workshops} ({VRW})}, pp. 920--921. IEEE Computer Society, Los Alamitos, CA, USA, Mar. 2022. doi: {{%
10\hspace{.1pt}\discretionary{.}{%
}{.}\hspace{.4pt}1109\discretionary{/}{%
}{/}VRW55335\hspace{.1pt}\discretionary{.}{%
}{.}\hspace{.4pt}2022\hspace{.1pt}\discretionary{.}{%
}{.}\hspace{.4pt}00311}}


\bibitem{mccrae_multiscale_2009}
J.~McCrae, I.~Mordatch, M.~Glueck, and A.~Khan.
\newblock Multiscale {3D} navigation.
\newblock In {\em Proceedings of the 2009 symposium on {Interactive} {3D} graphics and games}, pp. 7--14. ACM, Boston Massachusetts, Feb. 2009. doi: {{%
10\hspace{.1pt}\discretionary{.}{%
}{.}\hspace{.4pt}1145\discretionary{/}{%
}{/}1507149\hspace{.1pt}\discretionary{.}{%
}{.}\hspace{.4pt}1507151}}


\bibitem{mirhosseini_automatic_2017}
S.~Mirhosseini, I.~Gutenko, S.~Ojal, J.~Marino, and A.~E. Kaufman.
\newblock Automatic speed and direction control along constrained navigation paths.
\newblock In {\em 2017 {IEEE} {Virtual} {Reality} ({VR})}, pp. 29--36. IEEE, Los Angeles, CA, USA, 2017. doi: {{%
10\hspace{.1pt}\discretionary{.}{%
}{.}\hspace{.4pt}1109\discretionary{/}{%
}{/}VR\hspace{.1pt}\discretionary{.}{%
}{.}\hspace{.4pt}2017\hspace{.1pt}\discretionary{.}{%
}{.}\hspace{.4pt}7892228}}


\bibitem{pierce_navigation_2004}
J.~Pierce and R.~Pausch.
\newblock Navigation with place representations and visible landmarks.
\newblock In {\em {IEEE} {Virtual} {Reality} 2004}, pp. 173--288, 2004. doi: {{%
10\hspace{.1pt}\discretionary{.}{%
}{.}\hspace{.4pt}1109\discretionary{/}{%
}{/}VR\hspace{.1pt}\discretionary{.}{%
}{.}\hspace{.4pt}2004\hspace{.1pt}\discretionary{.}{%
}{.}\hspace{.4pt}1310071}}


\bibitem{piumsomboon_superman_2018}
T.~Piumsomboon, G.~A. Lee, B.~Ens, B.~H. Thomas, and M.~Billinghurst.
\newblock Superman vs {Giant}: {A} {Study} on {Spatial} {Perception} for a {Multi}-{Scale} {Mixed} {Reality} {Flying} {Telepresence} {Interface}.
\newblock {\em IEEE Transactions on Visualization and Computer Graphics}, 24(11):2974--2982, Nov. 2018. doi: {{%
10\hspace{.1pt}\discretionary{.}{%
}{.}\hspace{.4pt}1109\discretionary{/}{%
}{/}TVCG\hspace{.1pt}\discretionary{.}{%
}{.}\hspace{.4pt}2018\hspace{.1pt}\discretionary{.}{%
}{.}\hspace{.4pt}2868594}}


\bibitem{rampin_taguette_2021}
R.~Rampin and V.~Rampin.
\newblock Taguette: open-source qualitative data analysis.
\newblock {\em Journal of Open Source Software}, 6(68):3522, 2021. doi: {{%
10\hspace{.1pt}\discretionary{.}{%
}{.}\hspace{.4pt}21105\discretionary{/}{%
}{/}joss\hspace{.1pt}\discretionary{.}{%
}{.}\hspace{.4pt}03522}}


\bibitem{sousa_vrrrroom_2017}
M.~Sousa, D.~Mendes, S.~Paulo, N.~Matela, J.~Jorge, and D.~S. Lopes.
\newblock {VRRRRoom}: {Virtual} {Reality} for {Radiologists} in the {Reading} {Room}.
\newblock In {\em Proceedings of the 2017 {CHI} {Conference} on {Human} {Factors} in {Computing} {Systems}}, {CHI} '17, pp. 4057--4062. Association for Computing Machinery, New York, NY, USA, 2017. doi: {{%
10\hspace{.1pt}\discretionary{.}{%
}{.}\hspace{.4pt}1145\discretionary{/}{%
}{/}3025453\hspace{.1pt}\discretionary{.}{%
}{.}\hspace{.4pt}3025566}}


\bibitem{stoakley_virtual_1995}
R.~Stoakley, M.~J. Conway, and R.~Pausch.
\newblock Virtual reality on a {WIM}: interactive worlds in miniature.
\newblock In {\em Proceedings of the {SIGCHI} conference on {Human} factors in computing systems - {CHI} '95}, pp. 265--272. ACM Press, Denver, Colorado, United States, 1995. doi: {{%
10\hspace{.1pt}\discretionary{.}{%
}{.}\hspace{.4pt}1145\discretionary{/}{%
}{/}223904\hspace{.1pt}\discretionary{.}{%
}{.}\hspace{.4pt}223938}}


\bibitem{trindade_improving_2011}
D.~R. Trindade and A.~B. Raposo.
\newblock Improving {3D} navigation in multiscale environments using cubemap-based techniques.
\newblock In {\em Proceedings of the 2011 {ACM} {Symposium} on {Applied} {Computing}}, pp. 1215--1221. ACM, TaiChung Taiwan, Mar. 2011. doi: {{%
10\hspace{.1pt}\discretionary{.}{%
}{.}\hspace{.4pt}1145\discretionary{/}{%
}{/}1982185\hspace{.1pt}\discretionary{.}{%
}{.}\hspace{.4pt}1982454}}


\bibitem{viega_3d_1996}
J.~Viega, M.~J. Conway, G.~Williams, and R.~Pausch.
\newblock {3D} magic lenses.
\newblock In {\em Proceedings of the 9th {Annual} {ACM} {Symposium} on {User} {Interface} {Software} and {Technology}}, {UIST} '96, pp. 51--58. Association for Computing Machinery, New York, NY, USA, 1996. doi: {{%
10\hspace{.1pt}\discretionary{.}{%
}{.}\hspace{.4pt}1145\discretionary{/}{%
}{/}237091\hspace{.1pt}\discretionary{.}{%
}{.}\hspace{.4pt}237098}}


\bibitem{ware_context_1997}
C.~Ware and D.~Fleet.
\newblock Context sensitive flying interface.
\newblock In {\em Proceedings of the 1997 symposium on {Interactive} {3D} graphics - {SI3D} '97}, pp. 127--ff. ACM Press, Providence, Rhode Island, United States, 1997. doi: {{%
10\hspace{.1pt}\discretionary{.}{%
}{.}\hspace{.4pt}1145\discretionary{/}{%
}{/}253284\hspace{.1pt}\discretionary{.}{%
}{.}\hspace{.4pt}253319}}


\bibitem{wingrave_overcoming_2006}
C.~Wingrave, Y.~Haciahmetoglu, and D.~Bowman.
\newblock Overcoming {World} in {Miniature} {Limitations} by a {Scaled} and {Scrolling} {WIM}.
\newblock In {\em {3D} {User} {Interfaces} ({3DUI}'06)}, pp. 11--16. IEEE, Alexandria, VA, USA, 2006. doi: {{%
10\hspace{.1pt}\discretionary{.}{%
}{.}\hspace{.4pt}1109\discretionary{/}{%
}{/}VR\hspace{.1pt}\discretionary{.}{%
}{.}\hspace{.4pt}2006\hspace{.1pt}\discretionary{.}{%
}{.}\hspace{.4pt}106}}


\end{thebibliography}
\end{document}